\documentclass[
 reprint,
 superscriptaddress,
 amsmath,
 amssymb,
 aps,
 onecolumn,
 nofootinbib,
]{revtex4-1}
\usepackage{graphicx}
\usepackage{dcolumn}
\usepackage{bm}
\usepackage[caption=false]{subfig}
\usepackage{floatrow}
\usepackage{appendix}
\usepackage{amsmath}

\begin{document}
\captionsetup[figure]{labelfont={bf}}

\title{DC Hall coefficient of the strongly correlated Hubbard model}

\author{Wen O. Wang}
\email{wenwang.physics@gmail.com}
\affiliation{Department of Applied Physics, Stanford University, CA 94305, USA}

\author{Jixun K. Ding}
\affiliation{Department of Applied Physics, Stanford University, CA 94305, USA}

\author{Brian Moritz}
\affiliation{Stanford Institute for Materials and Energy Sciences,
SLAC National Accelerator Laboratory, 2575 Sand Hill Road, Menlo Park, CA 94025, USA}

\author{Edwin W. Huang}
\affiliation{Department of Physics and Institute of Condensed Matter Theory, University of Illinois at Urbana-Champaign, Urbana, IL 61801, USA}

\author{Thomas P. Devereaux}
\email{tpd@stanford.edu}
\affiliation{Stanford Institute for Materials and Energy Sciences,
SLAC National Accelerator Laboratory, 2575 Sand Hill Road, Menlo Park, CA 94025, USA}
\affiliation{
Department of Materials Science and Engineering, Stanford University, Stanford, CA 94305, USA}

\begin{abstract}
The Hall coefficient is related to the effective carrier density and Fermi surface topology in non-interacting and weakly interacting systems. In strongly correlated systems, the relation between the Hall coefficient and single-particle properties is less clear. Clarifying this relation would give insight into the nature of transport in strongly correlated materials that lack well-formed quasiparticles.
In this work, we investigate
the DC Hall coefficient of the Hubbard model
using determinant quantum Monte Carlo in conjunction with a recently developed expansion of magneto-transport coefficients in terms of thermodynamic susceptibilities. At leading order in the expansion, we observe a change of sign
in the Hall coefficient as a function of temperature and interaction strength, which we relate to a change in the topology of the 
apparent
Fermi surface. 
We also combine our Hall coefficient results with optical conductivity values 
to evaluate the Hall angle, as well as effective mobility  
and effective mass based on Drude theory of metals.
\end{abstract}

\date{\today}

\pacs{Valid PACS appear here}
\maketitle

\section*{Introduction}

The Hall coefficient $\mathit{R_\mathrm{H}}$ reveals properties of band structure and effective carrier density in weakly interacting systems, determined by the shape of the Fermi surface and the angular dependence of the quasiparticle relaxation time \cite{mermin,geometry}.
For strongly correlated materials, it may less directly correspond to the topology of the Fermi surface, since they generally lack well-formed quasiparticles.  Such materials exhibit unusual behaviors incompatible with the quasiparticle picture. Cuprates display large, $\mathit{T}$-linear resistivity \cite{linear-resistivity,linear-resistivity-2}, 
known as strange metallicity.
In some materials, 
magnetoresistance also shows unusual linear $\mathit{T}$-dependence \cite{magnetoresistance1,magnetoresistance2,magnetoresistance3}.  Recent experiments have shown that the Hall number may
be related closely to the strange metallicity \cite{strangemetalhighfieldnh}. 

$\mathit{R_\mathrm{H}}$ of high-$\mathit{T}_\mathrm{c}$ cuprates has strong temperature and doping dependence, in contrast to what is expected for free electrons. 
Underdoped cuprates have positive $\mathit{R_\mathrm{H}}$ with complicated temperature dependence \cite{ono2007strong}. As doping increases, $\mathit{R_\mathrm{H}}$ decreases and becomes $\mathit{T}$-independent at high temperature \cite{cuprateHall}. 
In the 
heavily 
overdoped regime, $\mathit{R_\mathrm{H}}$ experiences a sign change and becomes negative around $\mathit{p} = 0.3$ \cite{cupratehall2,negativehall}, 
in agreement with the doping dependent shape of the Fermi surface reported from angle-resolved photoemission spectroscopy (ARPES) \cite{ARPES,arpes2}. 
The doping dependence of $\mathit{R_\mathrm{H}}$ has been studied in several experiments \cite{badoux2016change,balakirev2003signature, strangemetalhighfieldnh, PhysRevB.95.224517}, and different theoretical models also have been established in order to explain this anomalous doping dependence of $\mathit{R_\mathrm{H}}$ \cite{verret2017phenomenological,storey2016hall,PhysRevLett.117.187001,PhysRevB.73.174501,charlebois2017hall}.
Finally, at low temperatures in the cuprates the cotangent of the Hall angle, $\mathit{\cot(\theta_\mathrm{H})}$, simply has quadratic temperature dependence
\cite{chien1991effect,cuprateHall,PhysRevLett.89.037003}.

Hubbard model calculations have revealed properties similar to those of high-$\mathit{T}_\mathrm{c}$ cuprates,
including $\mathit{T}$-linear resistivity in the strange metal phase \cite{huang}. 
Quantum Monte Carlo (QMC) simulations on the Hubbard model show similar generic nature of the quasiparticle  dispersion relation observed in some hole-doped cuprates, and demonstrate it to be mostly determined by the strong Coulomb repulsion, reflecting many-body correlations, rather than a simply one-electron band structure \cite{PhysRevB.50.7215}. 
Including a next-nearest neighbouring hopping $\mathit{t'}=-0.15$ for the Hubbard model ($\mathit{U}=8\mathit{t}$), they find the Fermi surface changes from a large hole-pocket centered at $(\pi,\pi)$ to an electron-pocket around $(0, 0)$ at $30\%$ doping. This implies the shape of the Fermi surface numerically measured in this model is in agreement with  the observed doping dependence of $\mathit{R_\mathrm{H}}$ in LSCO \cite{cupratehall2,negativehall}, if one assumes $\mathit{R_\mathrm{H}}$ is simply determined by the curvature of the Fermi surface.
 A change from a hole-like Fermi surface to an electron-like Fermi surface from low doping to high doping also has been observed for the Hubbard model with only nearest-neighbor hopping ($\mathit{t'}=0$) and strong interactions by other QMC simulations \cite{grober2000anomalous}, Dynamical cluster approximation (DCA) \cite{maier2002angle} techniques and a self-consistent projection
operator method (SCPM) \cite{PhysRevLett.94.156401}. 
Thus, we are motivated to calculate $\mathit{R_\mathrm{H}}$ in the Hubbard model to further investigate transport properties within the strange metal phase of cuprates.  Numerical calculations of $\mathit{R_\mathrm{H}}$ have been attempted for a number of models and with various algorithms, 
such as the 2D Hubbard model in the high frequency limit \cite{highf} and $\mathit{t}$-$\mathit{J}$ model with exact diagonalization \cite{tj2002}.
 In Ref. \cite{stanescu2003full}, it was demonstrated that $\mathit{R_\mathrm{H}}$ in a doped Mott insulator must change sign at $\mathit{p}<1/3$.
$\mathit{R_\mathrm{H}}$ at high temperature and high frequency has been examined in the $\mathit{t}$-$\mathit{J}$ model \cite{shastry1993faraday}, where they focused on the high frequency limit rather than the DC limit, 
because of the assumption that high-frequency $\mathit{R_\mathrm{H}^*\mathit}$ 
is instantaneous, and thus closer to the semiclassical expression $1/\mathit{n}^*\mathit{e}$.
However, in the Hubbard model the DC limit has been less well studied, especially using numerical techniques.

In this work, we calculate the DC Hall coefficient using an expansion that expresses magneto-transport coefficients in terms of a sum of thermodynamic susceptibilities \cite{assa,assa2}, avoiding challenges in numerical analytic continuation for obtaining DC transport properties.
We use the unbiased and numerically exact determinant quantum Monte Carlo (DQMC) algorithm \cite{dqmc1,dqmc2} to calculate the leading order term of the expansion of $\mathit{R_\mathrm{H}}$ from Ref.~\cite{assa}. We find strong temperature and doping dependence of $\mathit{R_\mathrm{H}}$ in a parameter regime with strong interactions and no coherent quasiparticles, and show a good correspondence between the sign of the Hall coefficient and the shape of a quasi-Fermi surface.

\section*{Results}
\subsection*{Hall Coefficient}
\begin{figure*}[htbp]
    \centering
    \includegraphics{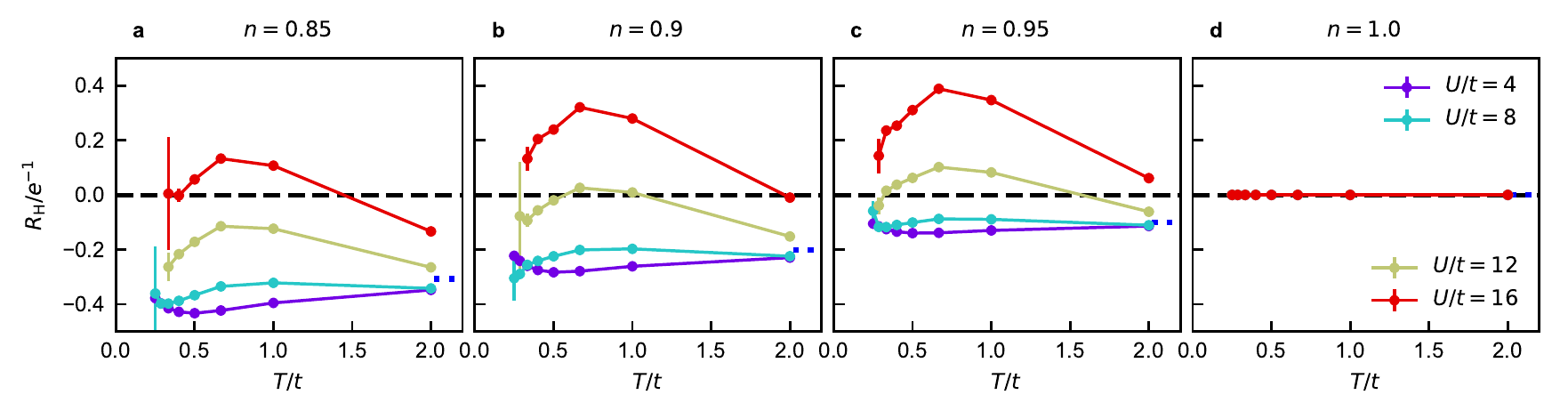} \caption{\label{fig:hall}
    \textbf{Hall coefficient.}
\textnormal{The Hall coefficient $\mathit{R_\mathrm{H}}$ obtained from DQMC for the Hubbard Model.
The simulations were performed on $8\times 8$ square lattice clusters, and coefficient evaluated as in Eq.\eqref{eq:rh}. 
$\mathit{n}$ is the charge density ((\textbf{a}-\textbf{d}) for $\mathit{n}=0.85,0.9,0.95,1.0$) and $\mathit{U}$ is the on-site Coulomb interaction in units of $\mathit{t}$.
$\mathit{R_\mathrm{H}}$ is given in units of $\mathit{e}^{-1}$ (lattice constant $\mathrm{a}=1$),
which is $\sim 1 \times 10^{-3}$ $\mathrm{cm}^3\mathrm{C}^{-1}$ for LSCO's lattice constants.
The blue dotted line marks the semi-classical estimate of $\mathit{R_\mathrm{H}}^{(0)}$ \cite{assa2}. }
}	
\end{figure*}

\begin{figure}[htbp]
    \centering
    \includegraphics{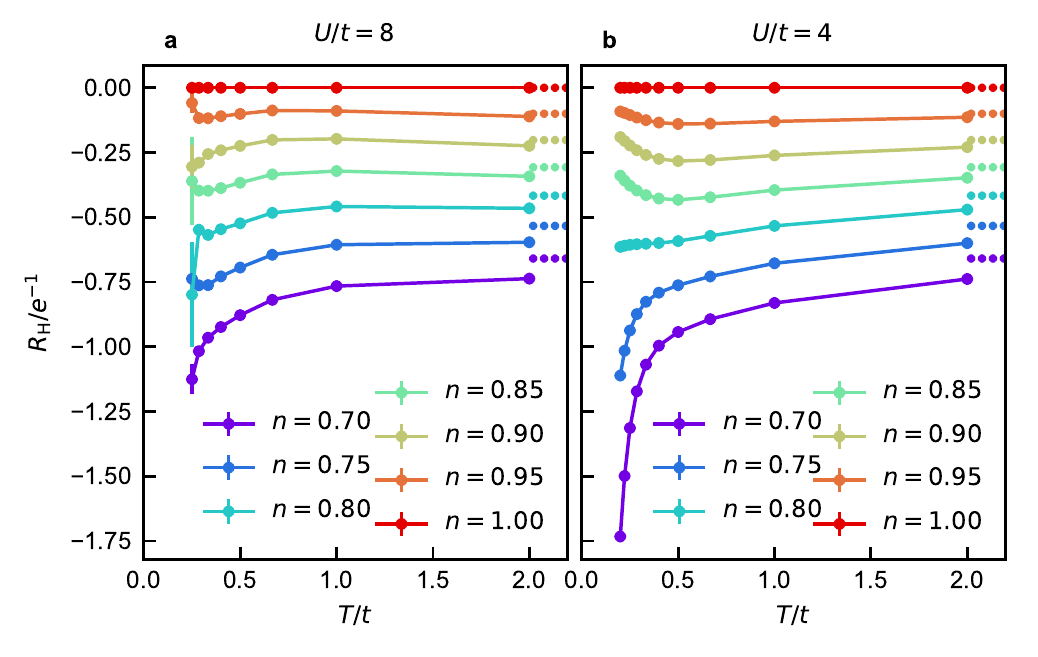}
    \caption{\label{fig:dop}
    \textbf{Hall coefficient with extended dopings.}
    \textnormal{The Hall coefficient $\mathit{R_\mathrm{H}}$ with extended dopings for 
    $\mathit{U/t}=8$
    (panel \textbf{a})
    and $\mathit{U/t}=4$
    (panel \textbf{b}).
    Data are obtained for $\mathit{U/t}=8$ (up to $\mathit{\beta}=3.5/\mathit{t}$) and $\mathit{U/t}=4$ (up to $\mathit{\beta}=5/\mathit{t}$) respectively, on a $8\times 8$ lattice.
    The dotted lines represent the semi-classical estimate of $\mathit{R_\mathrm{H}}^{(0)}$\cite{assa2}.
    }}
\end{figure}

In Fig.~\ref{fig:hall}, at half filling, particle-hole symmetry of the Hubbard Hamiltonian gives rise to a zero Hall coefficient for all values of $\mathit{U}$ as expected. 
As the system is doped away from half filling and the particle-hole symmetry is broken, $\mathit{R_\mathrm{H}}$ becomes nonzero and temperature dependent. When $\mathit{U}$ is small, the system is expected to be weakly interacting, and
the sign and magnitude of $\mathit{R_\mathrm{H}}$ is simply determined by the Fermi surface. Indeed, we see that for $\mathit{U}$ in the range between $4\mathit{t}$ and $8\mathit{t}$ in Fig.~\ref{fig:hall}, $\mathit{R_\mathrm{H}}$ has weak temperature dependence and is negative for all hole doping levels, corresponding to a well defined electron-like Fermi surface. For these same $\mathit{U}$ values in Fig.~\ref{fig:dop}, $\mathit{R_\mathrm{H}}$ has a nearly linear doping dependence, consistent with the quasiparticle picture and  Fig.~2 in Ref. \cite{assa2}.
With strong Coulomb interactions $\mathit{U}=12\mathit{t}$ and $16\mathit{t}$,  
we have $\mathit{T} \ll \mathit{U}$, and $\mathit{R_\mathrm{H}}$ becomes strongly temperature dependent and can be positive. 

\subsection*{Single-particle properties}
\begin{figure*}[htbp]
    \centering
\includegraphics{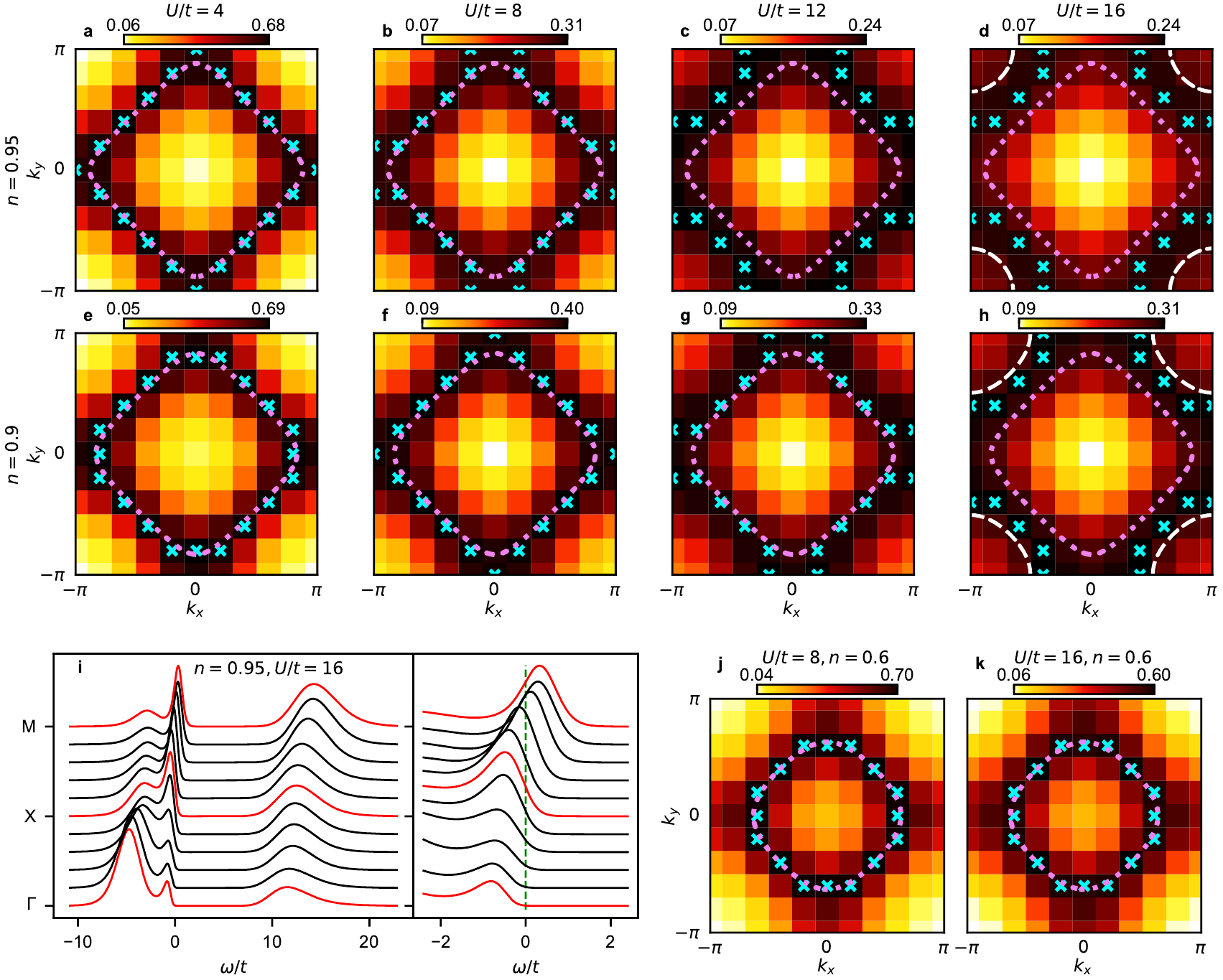}

\caption{
\textbf{Single-particle properties.}\textnormal{
\textbf{a-h} The imaginary time Green's function  $\mathit{G}(\mathbf{k},\mathit{\tau})\mathit{\beta}$ at $\mathit{\tau} = \mathit{\beta}/2$, in the first Brilliouin zone, as a proxy of the zero frequency spectral weights  $\mathit{A}(\mathbf{k},\mathit{\omega}=0)$.
The data is obtained from a $10\times 10$ lattice and at temperature $\mathit{T/t}=0.5$.
To roughly visualize the locus of intensity maxima, each momentum point with intensity greater than at least $6$ of its neighbors' is marked by a blue cross. ($4$ for the $X$ point if it has marked neighboring points).
The dashed lines for $\mathit{U/t}=16$ mark the Fermi surface in the Hubbard I approximation \cite{hubbard1963electron}.
The pink dotted lines are the Fermi surface for non interacting model.
\textbf{i} The spectral function $\mathit{A}(\mathbf{k},\mathit{\omega})$ along the high symmetry cuts $\Gamma$-$X$-$M$, with $\mathit{n}=0.95, \mathit{U/t}=16, \mathit{T/t}=0.5$, obtained via maximum entropy analytical continuation of $\mathit{G}(\mathbf{k},\mathit{\tau})$. 
Right panel shows a zoomed-in view of data near $\mathit{\omega}=0$.
\textbf{j-k} $\mathit{G}(\mathbf{k},\mathit{\tau}=\mathit{\beta}/2)\mathit{\beta}$ for $\mathit{n}=0.6$ and $\mathit{T/t}=0.5$, for $\mathit{U/t}=8$ and $\mathit{U/t}=16$ respectively. 
The blue crosses and pink dotted lines are as in \textbf{a-h}. 
}}
\label{fig:green}
\end{figure*}

To explore the connection between the Hall coefficient and quasi-Fermi surface in strongly interacting systems, we investigate the spectral weight around $\mathit{\omega} = 0$. 
$\mathit{G}(\mathbf{k},\mathit{\tau}=\mathit{\beta}/2)\mathit{\beta} $, as a proxy for $\mathit{A}(\mathbf{k}, \mathit{\omega}= 0)$ (see the "Methods" section), within the first Brillouin zone as shown in Figs.~\ref{fig:green}\textbf{a-h}. 
For weak interactions, the peak of $\mathit{G}(\mathbf{k},\mathit{\tau}=\mathit{\beta}/2)\mathit{\beta}$ in momentum space marks the position of the Fermi surface. 
For fixed hole doping, as the interaction gets stronger and opens a large Mott gap above the Fermi energy, $\mathit{R_\mathrm{H}}$ becomes positive and the peak of $\mathit{G}(\mathbf{k},\mathit{\tau}=\mathit{\beta}/2)\mathit{\beta}$ moves toward the $(\pi,\pi)$ point and the dashed lines, which mark the Fermi surface position predicted under the Hubbard-I approximation \cite{hubbard1963electron,grober2000anomalous}.
As $\mathit{U}$ becomes stronger, the Fermi surface changes from closed (a pocket centered at $\Gamma$ point) to open (a pocket centered at $M$ point). 
This evolution is shown for doping $\mathit{p} = 0.05$($\mathit{n}=0.95$) and $\mathit{p}=0.1$($\mathit{n}=0.9$).
Meanwhile, the spectral peak becomes broader, signaling that the Fermi surface becomes less well-defined as interaction strength increases.
However, we could still see a clear connection between $\mathit{R_\mathrm{H}}$ and the spectral weight, even without a well-defined Fermi surface or well-formed quasiparticles.
When the Fermi pocket changes from electron-like to hole-like, the sign of $\mathit{R_\mathrm{H}}$ changes from negative to positive [c.f. Fig.~\ref{fig:hall}].
For fixed Hubbard $U$, as doping level increases, the Fermi surface unsurprisingly moves back to $(0, 0)$ to enclose an electron pocket, as $\mathit{R_\mathrm{H}}$ decreases, returning to quasiparticle behavior. 
Within the low doping regime, the hole-like Fermi surface violates the Luttinger theorem, which is in agreement with other numerical results on the Hubbard model \cite{grober2000anomalous,maier2002angle,PhysRevLett.94.156401,sen2020mott,stanescu2004nonperturbative}.
The peak of $\mathit{G}(\mathbf{k},\mathit{\tau}=\mathit{\beta}/2)\mathit{\beta}$ becomes better defined going away from the Mott insulator, either by doping or decreasing $\mathit{U}$.
The evolution of the Fermi pocket is similar to ARPES experiments \cite{ARPES,arpes2}.
We also notice that for strong interactions as temperature decreases from $\mathit{T}=2\mathit{t}$ to $\mathit{T}\sim \mathit{t}/3$, we see that the peak of $\mathit{G}(\mathbf{k},\mathit{\tau}=\mathit{\beta}/2)\mathit{\beta}$ moves from close to $(0, 0)$ out towards $(\pi,\pi)$, and then moves slightly back towards $(0, 0)$, which can correspond to the two sign changes of $\mathit{R_\mathrm{H}}$ as a function of temperature in Fig. \ref{fig:hall}. 
We can see similar $\mathit{A}(\mathbf{k},\mathit{\omega})$ peak position changes in momentum space with temperature in a DMFT study \cite{deng2013bad}, and DQMC method accounts for momentum dependent self-energy effects.
Examples of  $\mathit{A}(\mathbf{k},\mathit{\omega})$ obtained from maximum entropy analytic continuation are shown in Fig.~\ref{fig:green}\textbf{i}.
Compared with  Fig.~\ref{fig:green}\textbf{d},
as we move along the $\Gamma$-$X$-$M$ momentum curve, the location of the spectral weight peak crosses $\mathit{\omega} = 0$ between $X$ and $M$,
indicating that our proxy $\mathit{G}(\mathbf{k}, \mathit{\beta}/2)$ properly represents the behavior of the spectral weight
and that the Fermi pocket is hole-like. 
Figs.~\ref{fig:green}\textbf{j-k} show the electron pocket  for both $\mathit{U/t}=8$ and $\mathit{U/t}=16$ at large hole-doping above $0.3$.
The Fermi surface positions are similar, 
and the spectral weight peaks are sharp, 
meaning that the coherence of $\mathit{A}(\mathbf{k},\mathit{\omega})$ with large doping is more consistent with a quasiparticle picture.
In contrast to $\mathit{n}=0.95$, at $\mathit{n}=0.6$ the apparent Fermi surface closely follows the non-interacting Fermi surface and is minimally affected by increasing interaction strength.

\subsection*{Hall Angle, Mobility and Mass}
For completeness, we also calculate
the Hall angle $\mathit{\cot(\theta_\mathrm{H})}$ and effective mass $\mathit{m}$ using $\mathit{R_\mathrm{H}}$ and $\mathit{\sigma_{xx}(\omega)}$ (see the "Methods" section), as shown in Fig.~\ref{fig:mm}. 
We observe a $\mathit{T^2}$ temperature dependence in $\mathit{\cot(\theta_\mathrm{H})}$ when temperature is low compared with the band width for most doping up to $\mathit{n}=0.9$ for $\mathit{U/t}=4$ and for temperatures higher than $\dfrac{1}{3.5}\mathit{t}$ for $\mathit{U/t}=8$, similar to what has been observed for LSCO \cite{cuprateHall,cupratehall2,PhysRevB56R8530} and other cuprates \cite{PhysRevB503246}.
For $\mathit{U/t} = 8$, the large error bars at the lowest temperature arise from a sever fermion sign problem \cite{PhysRevB.41.9301} which limits the accessible temperatures.
The upturn in $\mathit{\cot(\theta_\mathrm{H})}$ as temperature decreases for $\mathit{U}=4, \mathit{n}=0.95$ at the lowest temperatures,
probably results from anisotropy around the Fermi surface playing a much more significant role, 
considering it is relatively close to half filling \cite{PhysRevB.46.14297}.
When $\mathit{U}$ is strong ($\mathit{U/t}=8$ in Fig.~\ref{fig:mm}\textbf{c}) and doping is small, $\mathit{\cot(\theta_\mathrm{H})}$ shows a peak around $\mathit{T}\sim \mathit{t}$ (the ratio exceeds $1.0$).
Comparing this peak with the smooth $\mathit{\cot(\theta_\mathrm{H})}$ curve when $\mathit{U/t}=4$, we see again an indication that the Coulomb interaction strongly affects the temperature dependence of transport properties when $\mathit{T}\ll \mathit{U}$. 
The effective mass increases slightly as the temperature increases. We observe that a stronger interaction leads to a heavier effective mass. The mass approaches the mass of a free electron $\mathit{m}_\mathit{e}=\frac{1}{2\mathit{t}}$
at large doping and as the temperature tends to $0$, returning to a normal metal with well-defined quasiparticles as one would expect.

\section*{discussion}
In our results, we observe that when $\mathit{U}$ is large and doping is small, $\mathit{R_\mathrm{H}}$ in the Hubbard model exhibits complicated temperature and doping dependence.
Along with $\mathit{T}$-linear resistivity in the Hubbard model \cite{huang}, both phenomena suggest that strongly correlated electrons shouldn't simply behave like coherent quasiparticles moving in a static band structure. However, we also observe a corresondence between $\mathit{R_\mathrm{H}}$ and the topology of the Fermi surface, revealed by the proxy $\mathit{G}(\mathbf{k},\mathit{\beta}/2)\mathit{\beta}$. This is rather surprising, as the correspondence between $\mathit{R_\mathrm{H}}$ and Fermi surface topology is usually understood only in the quasiparticle picture for weakly interacting systems. Here, we have found this correspondence is still well established
even when strong correlations are present and the Fermi surface itself becomes ill-defined. 

The features of $\mathit{R_\mathrm{H}}$ are obtained from the single-band Hubbard model, using the unbiased and numerically exact DQMC algorithm. They directly show contributions to the Hall effect from the on-site Coulomb interaction and an
effective $\mathit{t'}$, 
pushing $\mathit{R_\mathrm{H}}$ to change sign and show strong temperature dependence and complicated doping dependence.
Comparing our $\mathit{R_\mathrm{H}}$ to that of cuprates \cite{cuprateHall,cupratehall2} at high temperatures,  such as LSCO, $\mathit{R_\mathrm{H}}$ usually changes sign  around $30\%$ hole doping. Underdoped cuprates at low temperature have complicated temperature dependence and almost unbounded Hall coefficient towards half filling.
Their low temperature behavior is affected jointly by the on-site Coulomb interaction and next nearest neighbour (NNN) hoping, as well as other experimental factors. However, our simulation corresponds to relatively high temperatures in LSCO experiments, before which unbounded $\mathit{R_\mathrm{H}}$  has alreay dropped down to the scale $\sim 10^{-3}\mathrm{cm}^3\mathrm{C}^{-1}$. Nevertheless, around the  point at which the sign changes, the order of magnitude of the ratio $\mathit{\delta R_\mathrm{H}}/\mathit{\delta p}$ in our $\mathit{R_\mathrm{H}}$ data in the Hubbard model is comparable to that of LSCO \cite{cuprateHall,cupratehall2, negativehall} at high temperatures.
Furthermore, here we have only focused on the  single-band Hubbard model with only nearest-neighbor hopping. 
The next-nearest-neighbor hoping can also deform the Fermi surface \cite{duffy1995influence} and affect $\mathit{R_\mathrm{H}}$.
Thus far, we have only implemented  the lowest order term of the effective expansion from Ref.  \cite{assa}. Correction terms involve tens of thousands of Wick contractions and are not feasible to simulate given current computational capacity. However, our results regarding sign changes using the leading order term are consistent with various other methods, including coupling the Hamiltonian to an external magnetic field (Ding, J.~K. et al. Manuscript in
preparation).

\section*{methods}
\subsection*{Hall Coefficient}
We calculate the Hall coefficient $\mathit{R_\mathrm{H}}$ in the doped Hubbard Model on a 2D square lattice with periodic boundary conditions, defined by the Hamiltonian
\begin{align*}
\mathit{H} = -\mathit{t}\sum_{\langle \mathit{jk}\rangle
,\sigma} \mathit{c_{j,\sigma}^{\dagger} c_{k,\sigma}}
 \mathrm{e}^{\mathrm{i} \int_\mathit{j}^\mathit{k} \mathit{e}\mathbf{A}(\mathbf{r})  d\mathbf{r}} 
- \mu\sum_{\mathit{j},\mathit{\sigma}} \mathit{n}_{\mathit{j},\mathit{\sigma}} + \mathit{U}\sum_{\mathit{j}}\mathit{n}_{\mathit{j},\uparrow
} \mathit{n}_{\mathit{j},\downarrow}
\stepcounter{equation}\tag{\theequation}\label{eq:hubbard}
\end{align*}
where $\mathit{t}$ is nearest-neighbor hopping energy, $\mathit{\mu}$ is chemical potential and $\mathit{U}$ is the Coulomb interaction. $\mathit{c}_{\mathit{j},\mathit{\sigma}}^{\dagger}$ stands for the creation operator for an electron on site $\mathit{j}$ with spin $\mathit{\sigma}$. $\mathit{n}_{\mathit{j},\mathit{\sigma}} \equiv \mathit{c}_{\mathit{j},\mathit{\sigma}}^{\dagger} \mathit{c}_{\mathit{j},\mathit{\sigma}}$ is the number operator.
$\mathit{\theta}_{\mathit{jk}} = \int_\mathit{j}^\mathit{k} \mathit{e}\mathbf{A}(\mathbf{r})d\mathbf{r}$ is the Peierls phase factor. For a perpendicular field $\mathbf{B} = \mathit{B} \hat{z}$, we choose the vector potential $\mathbf{A}= -\mathit{\alpha B y\hat{x} + (\mathrm{1}-\alpha) B x\hat{y}}$, with $\mathit{\alpha}$ associated with an arbitrary gauge choice.

The DC Hall coefficient $\mathit{R_\mathrm{H}}$ \cite{assa,assa2} is expressed as 
\begin{align*}
&\mathit{R_\mathrm{H}}^{(0)} = -\operatorname{Im}\frac{\mathit{e}^2\mathit{t}^2/\mathit{V}}{ (\int^{\mathit{\beta}}_0 d\mathit{\tau} \langle \mathit{j_x(\tau)j_x\rangle/V})^2 }\int^{\mathit{\beta}}_0 d\mathit{\tau}[-(1-\mathit{\alpha})\times\\
& \langle \mathit{j_y(\tau)} \sum_{\mathit{k},\mathit{\sigma}}(\mathit{c}_{\mathit{k}+\mathit{\delta\hat{x}},\mathit{\sigma}}^\dagger \mathit{c}_{\mathit{k}+\mathit{\delta\hat{y}},\mathit{\sigma}}+\mathit{c}_{\mathit{k},\mathit{\sigma}}^\dagger \mathit{c}_{\mathit{k}+\mathit{\delta\hat{x}}+\mathit{\delta\hat{y}},\mathit{\sigma}} - \mathrm{h.c.})\rangle \\
&+\mathit{\alpha} \langle \mathit{j_x(\tau)\sum_{k,\sigma}(c_{k+\delta\hat{x}+\delta\hat{y},\sigma}^\dagger c_{k,\sigma}+c_{k+\delta\hat{x},\sigma}^\dagger c_{k+\delta\hat{y},\sigma}} - \mathrm{h.c.})\rangle\stepcounter{equation}\tag{\theequation}\label{eq:rh}
\end{align*}
where $\mathit{j_x}$ and $\mathit{j_y}$ are current operators along $\mathit{x}$ and $\mathit{y}$ directions.
For example, $\mathit{j_x} =-\mathrm{i}\mathit{e}t \sum_{\mathit{k},\mathit{\sigma}}(\mathit{c_{k+\delta\hat{x},\sigma}^\dagger c_{k,\sigma}}  - \mathrm{h.c.})$.
By $C_4$ rotational symmetry, we notice that the magnitude of the term after $1-\mathit{\alpha}$ is equal to the term after $\alpha$, leaving the expression independent of $\mathit{\alpha}$ and gauge invariant. 

We use DQMC to calculate the susceptibilities in Eq.~\eqref{eq:rh} to obtain $\mathit{R_\mathrm{H}}^{(0)}$ (shown in Figs. \ref{fig:hall}, \ref{fig:dop}).
We measure both unequal time correlators in  Eq.~\eqref{eq:rh} and combine them by selecting $\alpha=0.5$, as in Refs. \cite{assa,assa2}.
Due to the fermion sign problem, a large number of measurements is required to cope with the small sign, which limits the temperatures we can access. Nevertheless, we can access temperatures below the spin exchange energy $\mathit{J}=4\mathit{t}^2/\mathit{U}$ reliably for all doping levels.
The finite size effect is minimal in our results (Supplementary Fig. 1).

Limitations of our method for evaluating $\mathit{R_\mathrm{H}}$ include: (1) The fermion sign problem, which constrains our ability to access lower temperatures. (2) Correction terms of the effective expansion involve a proliferation of Wick contractions and are not implemented given current computational capacity. (3) The next-nearest-neighbour hoping has not been taken into account.

\subsection*{Single-Particle Properties}
The spectral function $\mathit{A}(\mathbf{k},\mathit{\omega})$ on all frequencies can be computed by adopting standard maximum entropy analytic continuation \cite{ana1,ana2}. Starting from the imaginary time Green's function data $\mathit{G}(\mathbf{k},\mathit{\tau}) = \langle \mathit{c}(\mathbf{k},\mathit{\tau})\mathit{c}^\dagger(\mathbf{k},0)\rangle$,
we invert the relation
\begin{equation}
    \mathit{G}(\mathbf{k},\mathit{\tau})=\int^{\infty}_{-\infty}d\mathit{\omega} \frac{ \mathrm{e}^{-\mathit{\tau} \mathit{\omega}}}{1 + \mathrm{e}^{-\mathit{\beta\omega}}} \mathit{A}(\mathbf{k},\mathit{\omega}).
\end{equation}
We also calculate a proxy for $\mathit{A}(\mathbf{k},\mathit{\omega}=0)$, showing the position of the Fermi surface without the need for analytic continuation.
$\mathit{A}(\mathbf{k},\mathit{\omega}=0)$ can be approximately calculated directly as $\mathit{G}(\mathbf{k},\mathit{\tau} = \mathit{\beta} /2)\mathit{\beta}$ (Fig.~\ref{fig:green}), 
since $\mathit{\tau}=\mathit{\beta}/2$ contains the largest weight of $\mathit{A}(\mathbf{k},\mathit{\omega}) = -\dfrac{1}{\pi}\operatorname{Im}\mathit{G}(\mathbf{k},\mathit{\omega})$ near $\mathit{\omega} = 0$. We see this from the relation
\begin{align*}
\mathit{G}(\mathbf{k},\mathit{\tau}=\mathit{\beta}/2) &= \langle \mathit{c}_{\mathbf{k}}(\mathit{\tau}=\mathit{\beta}/2)\mathit{c}_{\mathbf{k}}^\dagger \rangle =- \int \frac{d\mathit{\omega}}{\pi} \frac{1}{2\cosh(\mathit{\beta\omega}/2)} \operatorname{Im} \mathit{G}(\mathbf{k},\mathit{\omega}).
\end{align*} 

\subsection*{Hall Angle and Mass}
The Hall angle $\mathit{\theta_\mathrm{H}}$ is defined by $\cot\mathit{\theta_\mathrm{H}} = \mathit{\sigma}_{\mathit{xx}}/\mathit{\sigma_{xy}}$. So from $\mathit{R_\mathrm{H}} \Big|_{\mathit{B}=0} = \mathit{\sigma_{xy}\big{/}{\sigma^\mathrm{2}_{xx} B}}\Big|_{\mathit{B}=0}$ and DC optical conductivity $\mathit{\sigma_{xx}}$, we can evaluate the Hall angle with
\begin{equation}
    \cot(\mathit{\theta_\mathrm{H}})\mathit{B}\big|_{\mathit{B}=0} = \frac{1}{\mathit{R_\mathrm{H}} \mathit{\sigma_{xx}}}\bigg|_{\mathit{B}=0}.
\end{equation}

\begin{figure*}[htbp]
    \centering
    
    \includegraphics{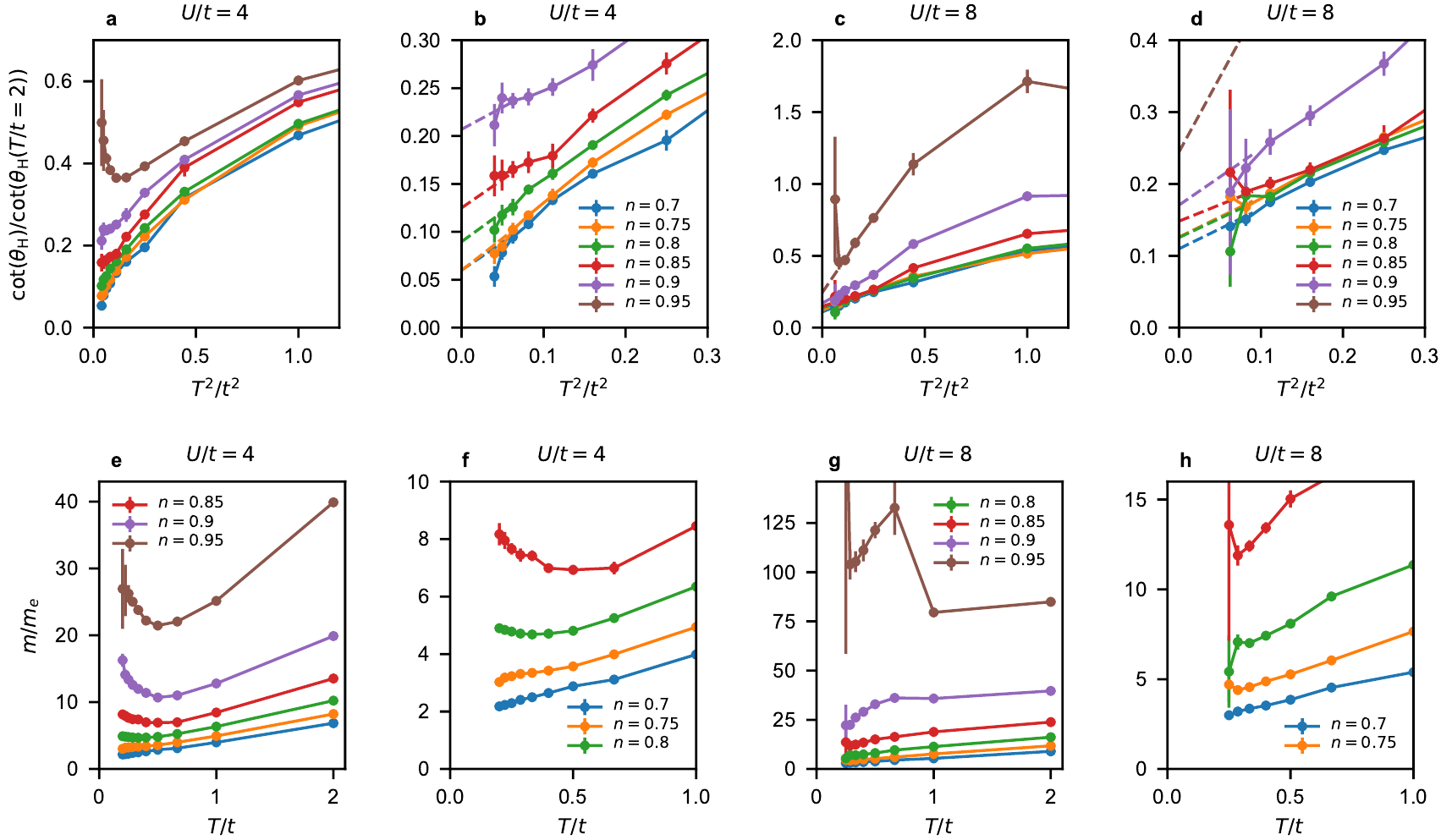}
   
\caption{
\textbf{Hall angle and mass.}
\textnormal{
\textbf{a-d} The Hall angle $\mathit{\theta_\mathrm{H}}$ obtained from DQMC, normalized and shown as $\cot(\mathit{\theta_\mathrm{H}})/\cot(\mathit{\theta_\mathrm{H}}(\mathit{T/t}=2))$ for $\mathit{U/t}=4$ and $\mathit{U/t}=8$, with zoomed in versions of each plot on the right. Dashed lines are a guide to eyes. 
Calculations are done on a $8\times 8$ lattice. 
\textbf{e-h} The effective mass obtained from DQMC. 
The unit is $\frac{1}{2\mathit{t}}= \mathit{m}_\mathit{e}$, where $\mathit{m}_\mathit{e}$ is the effective mass of a free electron in a non-interacting tight binding system.
Calculations are done on a $8\times 8$ lattice.
For panels \textbf{c}-\textbf{d} and \textbf{g}-\textbf{h}, the error bars at the lowest temperature ($\mathit{T/t} = 0.25$)
for $\mathit{n}=0.95$ is reduced by a factor of $50$ and for $\mathit{n}=0.85$ is reduced by a factor of $15$ to be shown to avoid overlapping.
}
}
\label{fig:mm}
\end{figure*}

Under the assumption of a single quasiparticle Fermi pocket, 
we can use the Drude theory of metals to
write $\mathit{R_\mathrm{H}} = 1/(\mathit{n}^{*}\mathit{e})$ and $\mathit{\sigma_{xx}} = \mathit{n}^{*}\mathit{e}\mathit{\mu}$, where $\mathit{\mu}$ is the effective mobility with a convention that $n^*$ is negative for electrons and positive for holes, so that
mobility is simply
\begin{equation}
    \mathit{\mu} = \mathit{\sigma_{xx}} \times \mathit{R_\mathrm{H}}
\end{equation} 
which itself is related to 
the Hall angle by $\cot(\mathit{\theta_\mathrm{H}})\mathit{B}\big|_{\mathit{B}=0} = 1/\mathit{\mu}$.  The optical conductivity $\mathit{\sigma_{xx}(\omega)}$ of the Hubbard Model has been investigated already with DQMC and maximum entropy analytic continuation \cite{huang}, whose methods we adapt here.
With relaxation time $\mathit{\tau}$ obtained from the inverse width of the Drude peak of $\mathit{\sigma_{xx}(\omega)}$, the effective mass of carriers (Figs.~\ref{fig:mm}\textbf{e-h}) could be evaluated under Drude theory using $\mathit{\sigma_{xx}} = -\dfrac{\mathit{n}^{*}\mathit{e}^2\mathit{\tau}}{\mathit{m}}$.
Thus we have the expression 
\begin{equation}
    \mathit{m} =- \frac{\mathit{\tau} \mathit{e}}{\mathit{R_\mathrm{H}} \mathit{\sigma_{xx}}}.
\end{equation}

There are different ways to determine the relaxation time $\mathit{\tau}$ (or frequency $\mathit{\omega_\tau}$) from $\mathit{\sigma_{xx}(\omega)}$. Here we choose the frequency $\mathit{\omega_\tau}$ where $\mathit{\sigma_{xx}(\omega_\tau)} = \mathit{\sigma_{xx}(\omega=\mathrm{0})}/2$. A special point in Fig.~\ref{fig:mm}\textbf{g} is $\mathit{U/t}=8, \mathit{n}=0.95, \mathit{T/t}=1$. For these parameters, $\sigma_{xx}(\omega)$ has a significant high frequency peak centered around $\mathit{ \omega \sim U}$, so the Drude peak does not decay to half of its zero frequency value before increasing again \cite{huang}. For these parameters, we select $\mathit{\omega_\tau}$ as the local minimum of $\mathit{\sigma_{xx}(\omega)}$ between the zero frequency Drude peak and the high-frequency peak at around $\mathit{ \omega \sim U}$, where the ratio at the minimum is  $\mathit{\sigma_{xx}(\omega_\tau)}/ \mathit{\sigma_{xx}(\omega=\mathrm{0})} = 0.655$. We also can fit the frequency dependence of $\mathit{\sigma_{xx}(\omega)}$ to a zero frequency Lorentzian and a high-frequency Lorentzian or Gaussian, which yield $1.04\mathit{\tau}_0$ (Lorentzian) and  $0.91\mathit{\tau}_0$ (Gaussian), where $\mathit{\tau}_0$ is the value obtained from the local minimum method.
Using these different methods only changes 
the effective mass result slightly, 
but does not affect the features in Figs.~\ref{fig:mm}\textbf{e-h}.

\section*{Error analysis}
For our Hall coefficient results, we use jackknife resampling to calculate standard errors. Error bars represent $1$ standard error.
Error bars for measurements involving $\mathit{\sigma_{xx}(\omega)}$ represent random sampling errors, determined
by bootstrap resampling standard deviation \cite{huang}. Error bars represent $1$ bootstrap standard error.

\section*{Data availability}
Data supporting this manuscript are stored on the Sherlock cluster at Stanford University and are available from the corresponding author upon request. 

\section*{Code availability}
Source code for the simulations can be found at 
\url{https://doi.org/10.5281/zenodo.3923215}.

\section*{Acknowledgements}
We acknowledge helpful discussions with A. Auerbach, I. Khait, D. Scalapino, E. Berg, Y. Schattner, S. Kivelson and X.X. Huang. 
\emph{Funding:}
This work was supported by the U.S. Department of Energy (DOE), Office of Basic Energy Sciences,
Division of Materials Sciences and Engineering. 
EWH was supported by the Gordon and Betty Moore Foundation EPiQS Initiative through the grant GBMF 4305.
Computational work was performed on the Sherlock cluster at Stanford University and on resources of the National Energy Research Scientific Computing Center, supported by the U.S. DOE
under Contract no. DE-AC02-05CH11231.

\section*{Author contributions}
WOW performed numerical simulations and analyzed data.
EWH and TPD conceived the project.
All authors assisted in data interpretation and contributed to writing the manuscript. 
\section*{Competing interests}
 The authors declare no competing interest.

\clearpage
\renewcommand{\figurename}{\bfseries Fig.}

\floatsetup[figure]{style=plain, subcapbesideposition=top}

\clearpage
\newpage

\clearpage
\renewcommand{\figurename}{\bfseries Supplementary Fig.}
\floatsetup[figure]{style=plain, subcapbesideposition=top}
\setcounter{figure}{0}

\begin{figure}[htbp]
    \centering
    \includegraphics{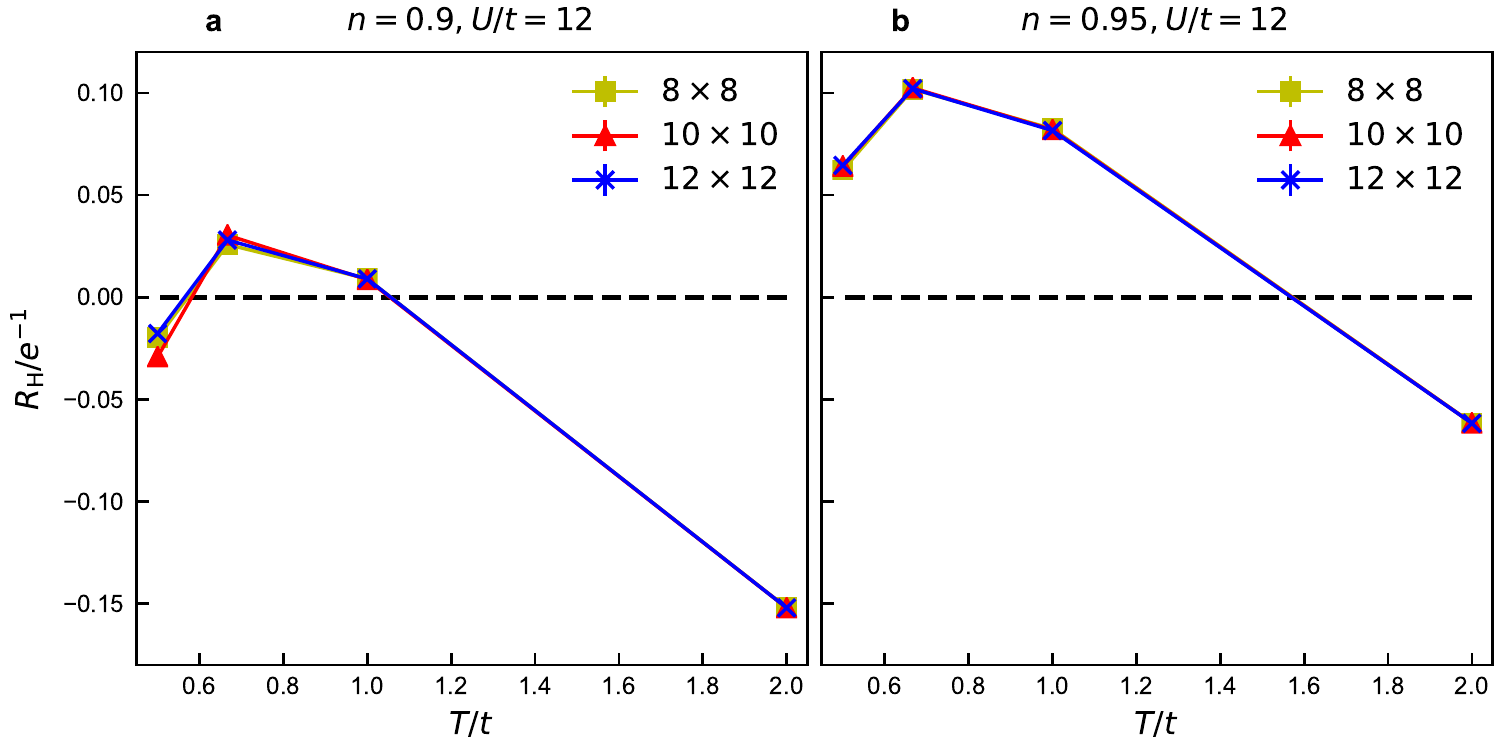}
    \caption{\label{fig:finitesize}
    \textbf{Finite size analysis of Hall coefficient.}
    \textnormal{Finite size analysis for the Hall coefficient $\mathit{R_\mathrm{H}}$ for $\mathit{U}/\mathit{t}=12, \mathit{n}=0.9$ (panel \textbf{a}) and $\mathit{U}/\mathit{t}=12, \mathit{n}=0.95$ (panel \textbf{b}) for lattice sizes $8\times 8$, $10\times 10$, and $12\times 12$.
    We can conclude the finite size effects are minimal in our results.
    }}
\end{figure}
\end{document}